\documentclass[10pt]{article}
\usepackage[latin1]{inputenc}
\usepackage{amsmath}
\usepackage{amsfonts}
\usepackage{amssymb}
\usepackage{graphicx}
\usepackage{epstopdf}
\usepackage{color}
\usepackage{appendix}
\usepackage{csvsimple}
\usepackage{longtable}
\usepackage{chngpage}
\usepackage{subcaption}
\usepackage[export]{adjustbox}[2011/08/13]
\usepackage{mathrsfs}
\usepackage{a4wide}
\usepackage{psfrag}
\usepackage{multirow}
\usepackage{hhline}

\begin{document}

\title{\bf Modelling Inflation with a Power-law Approach to the Inflationary Plateau}

\author{Konstantinos Dimopoulos and Charlotte Owen\\
\\
{\small\em Consortium for Fundamental Physics, Physics Department,}\\
{\small\em Lancaster University, Lancaster LA1 4YB, UK}
\\
\vspace{-.4cm}
\\
{\small e-mails: {\tt k.dimopoulos1@lancaster.ac.uk}, \
{\tt c.owen@lancaster.ac.uk}}}
\vspace{1cm}

\maketitle

\begin{abstract}
A new family of inflationary models is introduced and analysed. The behaviour 
of the parameters characterising the models suggest preferred values, which 
generate the most interesting testable predictions. Results are further improved
if late reheating and/or a subsequent period of thermal inflation is taken into
account. Specific model realisations consider a sub-Planckian inflaton variation
or a potential without fine-tuning of mass scales, based on the Planck and grand unified theory 
scales. A toy model realisation in the context of global and local supersymmetry
is examined and results fitting the Planck observations are determined. 
\end{abstract}

\nopagebreak

\section{Introduction}\leavevmode\\
Different classes of inflationary models produce wildly varying predictions for
observables with some models being %almost entirely 
ruled out by recent advances in the precision of cosmic microwave background observations. The Planck 
2015 results~\cite{Ade:2015lrj} found the following value for the spectral 
index of the curvature perturbation, $n_s$, and the following upper 
bound on the ratio of the power spectra of the tensor to scalar perturbations, 
$r$: $n_{s} = 0.968 \pm 0.006$ at 1-$\sigma$ with negligible tensors and 
$r<0.11$. 
These (and the fact that non-Gaussian and isocurvature perturbations have not 
been observed) strongly suggest that primordial inflation is single-field with 
a concave scalar potential featuring an inflationary plateau.

Indeed, while the minimal versions of chaotic and hybrid inflation are all but 
ruled out (but see Refs.~\cite{dihybrid,apostolos}), models with such an 
inflationary plateau (e.g. 
Starobinsky $R^2$ Inflation \cite{R2}, or Higgs Inflation \cite{Higgs})
have received enormous attention. In these models (and similar other such 
models, e.g. $\alpha$-attractors \cite{alpha}), the approach to the inflationary
plateau is exponential. In this case, however, distinguishing between models is 
difficult \cite{allthesame}. In contrast, a power-law inflationary plateau was 
considered in Ref.~\cite{shaft}, where shaft inflation was introduced, based
in global supersymmetry with a deceptively simple but highly non-perturbative
superpotential \mbox{$W\propto(\Phi^n+m^n)^{1/n}$}, where $m$ is a mass scale.

In this paper, we consider a new family of single-field inflationary models, 
which feature a power-law approach to the inflationary plateau but are not based
on an exotic superpotential like in Shaft Inflation. We call this family of 
models power-law plateau inflation.%
\footnote{This should not be confused with plateau inflation in Ref.\cite{plateau}.}
Power-law plateau inflation is characterised by a simple 
two-mass-scale potential, which, for large values of the inflaton field,
features the inflationary plateau but for small values, after the end of 
inflation, the potential is approximately monomial. The family is parameterised
by two real parameters, whose optimal values are determined by contrasting the 
model with observations. We find that the predicted spectral index of the 
curvature perturbation is near the sweet spot of the 
%falls comfortably inside the 1-$\sigma$ bounds of
Planck observations. For a sub-Planckian inflaton, the Lyth bound does not 
allow for a large value of the tensor to scalar ratio $r$. However, for mildly 
super-Planckian inflaton values we obtain sizeable $r$, which is easily testable
in the near future.

Even though our treatment is phenomenological, and the form of the potential
of power-law plateau inflation is data driven, we develop a simple toy model in global 
and local supersymmetry to demonstrate how power-law plateau inflation may well be 
realised in the context of fundamental theory.

We use natural units, where $c=\hbar=1$ and Newton's gravitational constant is 
$8\pi G=m_{P}^{-2}$, with $m_{P}=2.43\times 10^{18}\mathrm{GeV}$ being the reduced
Planck mass. 

\section{Power-Law Plateau Inflation}\leavevmode\\
We start from the following proposed potential:

\begin{equation}
V = V_{0}\Big(\frac{\varphi^{n}}{\varphi^{n} + m^{n}}\Big)^{q}\,,
\label{eq:main}
\end{equation}
where $m$ is a mass scale, $n$ and $q$ are real parameters and $\varphi$ is a 
canonically normalised, real scalar field. $V_{0}$ is a constant density scale
and we assume $\varphi \geq m$ because otherwise our model is indistinguishable
from monomial inflation with $V \propto \varphi^{nq}$, which is disfavoured by
observations. Initially we impose sub-Planckian values of $\varphi$ to compare 
with perturbative models and avoid supergravity (SUGRA) corrections. 

For the slow roll parameters, to first order in 
%$(\frac{m}{\varphi})^{n}$, 
$(m/\varphi)^{n}$, we find:
\begin{equation}
\epsilon = \frac{m_{p}^{2}}{2}\Big(\frac{V'}{V}\Big)^{2} \simeq 
\frac{n^{2}q^{2}}{2\alpha^{2}}\Big(\frac{m}{\varphi}\Big)^{2(n+1)}
\Big[1-2\Big(\frac{m}{\varphi}\Big)^{n}\Big]\,,
\label{eq:firstorderepsiloncontainingphi}
\end{equation}
\begin{equation}
\eta = m_{p}^{2}\Big(\frac{V''}{V}\Big) \simeq 
-\frac{n(n+1)q}{\alpha^{2}}\Big(\frac{m}{\varphi}\Big)^{n+2}
\Big[1-\frac{(nq+2n+1)}{(n+1)}\Big(\frac{m}{\varphi}\Big)^{n}\Big]\,,
\label{eq:firstorderetacontainingphi}
\end{equation}
where the prime denotes a derivative with respect to the scalar field and the 
second term in the square brackets is the first order correction. Considering
this correction allows us to approach the bound $\varphi\simeq m$. In the above
and throughout, $\alpha$ is defined as:
\begin{equation}
\alpha \equiv \frac{m}{m_{P}}.
\label{alpha}
\end{equation}
These result in the following expressions for the tensor to scalar ratio 
and the spectral index of the curvature perturbation:
\begin{equation}
r = 16\epsilon \simeq \frac{8n^{2}q^{2}}{\alpha^{2}}
\Big(\frac{m}{\varphi}\Big)^{2(n+1)}\Big[1-2\Big(\frac{m}{\varphi}\Big)^{n}\Big]\,,
\label{eq:firstorderrwithphi}
\end{equation}
\begin{equation}
n_{s} = 1+ 2\eta -6\epsilon \simeq 1 - \frac{2n(n+1)q}{\alpha^2}
\Big(\frac{m}{\varphi}\Big)^{n+2}\Big[1-\frac{(nq+2n+1)}{n+1}
\Big(\frac{m}{\varphi}\Big)^{n}\Big]\,,
\label{eq:firstordernswithphi}
\end{equation}
where the $\epsilon$ term has been omitted in Eq.\eqref{eq:firstordernswithphi}
because it is negligible. We can express:
\begin{equation}
%\Big(
\frac{\varphi}{m}
%\Big) 
\approx\Big[n(n+2)q\alpha^{-2} \Big(N + \frac{n+1}{n+2}\Big)\Big]^{\frac{1}{n+2}} 
\Big\{1-\frac{1}{2}\Big[n(n+2)q\alpha^{-2}
\Big(N+\frac{n+1}{n+2}\Big)\Big]^{-\frac{n}{n+2}}\Big\} \,,
\label{eq:firstorderphioverm}
\end{equation}
where we found the critical value $\varphi_{e}$, which ends inflation,
by setting $\eta$ to 1 in Eq.~\eqref{eq:firstorderepsiloncontainingphi}.%
\footnote{Again, the second term in the curly brackets is the first order 
correction, which allows us to approach $\varphi\simeq m$.} 
Therefore, we can write Eqs.~\eqref{eq:firstorderr} and \eqref{eq:firstorderns}
in terms of the remaining e-folds of inflation, $N$, as:
\begin{equation}
r=8n^{2}q^{2}\alpha^{\frac{2n}{n+2}}
\Big[n(n+2)q\Big(N + \frac{n+1}{n+2}\Big)\Big]^{-2\frac{n+1}{n+2}}
\Big\{1+(n-1)\Big[n(n+2)q\alpha^{-2}
\Big(N+\frac{n+1}{n+2}\Big)\Big]^{-\frac{n}{n+2}}\Big\}\,,
\label{eq:firstorderr}
\end{equation}
and
\begin{equation}
n_{s} = 1 - 2\frac{n+1}{n+2}\Big(N + \frac{n+1}{n+2}\Big)^{-1}
\Big\{1 + \frac{(n^{2}-2nq-n)}{2(n+1)} 
\Big[\frac{n(n+2)q}{\alpha^{2}} 
\Big(N + \frac{n+1}{n+2}\Big)\Big]^{-\frac{n}{n+2}}\Big\}.
\label{eq:firstorderns}
\end{equation}

\begin{figure}[h]
\centering
\includegraphics[width=\textwidth]{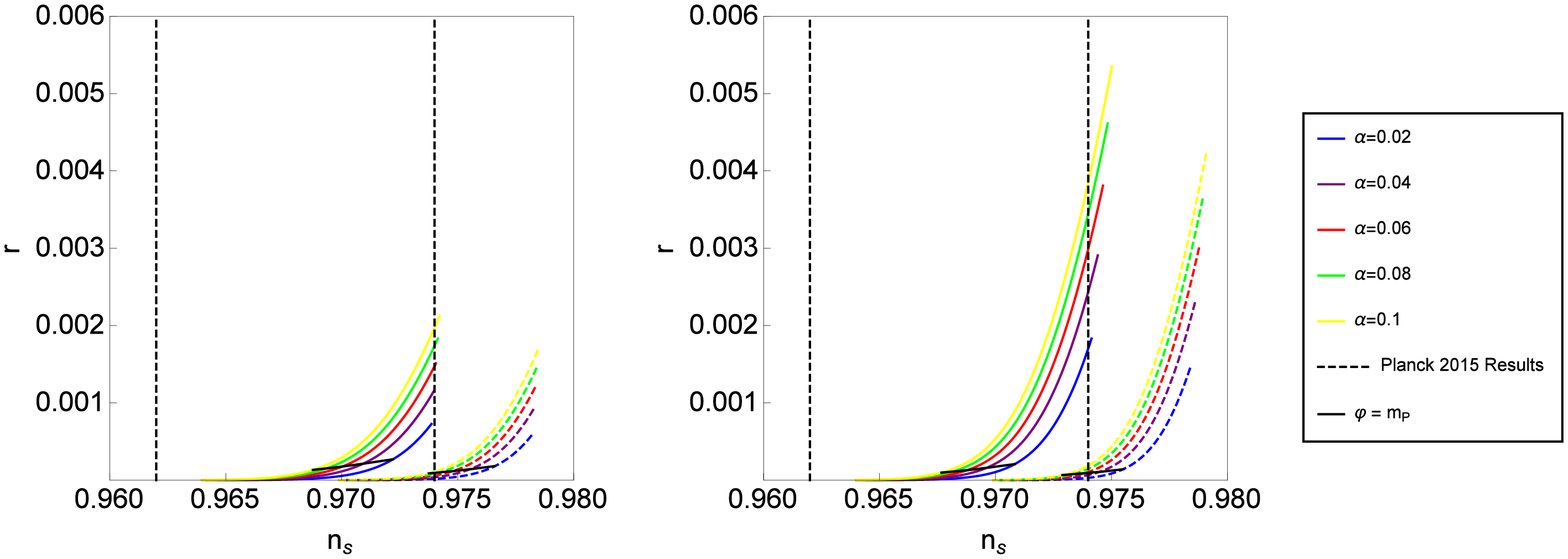}
\caption[]{%
The predictions of power-law plateau inflation in a graph depicting the tensor to scalar 
ratio $r$ in terms of the spectral index $n_s$.
Solid (dashed) lines correspond to $N=50$ ($N=60$) respectively. On 
the left we have $q=1$ and on the right $q=4$. $n$ increases along the length 
of the lines, counter-intuitively, \textit{right to left}. The vertical black 
dashed straight lines represent the 1-$\sigma$ bounds on $n_s$ from the Planck 
data. The slanted solid black line depicts the limit below which values of 
$\varphi$ are sub-Planckian.}
\label{fig:AllPlots}
\end{figure}

\begin{figure}[h]
\centering
\includegraphics[width=\textwidth]{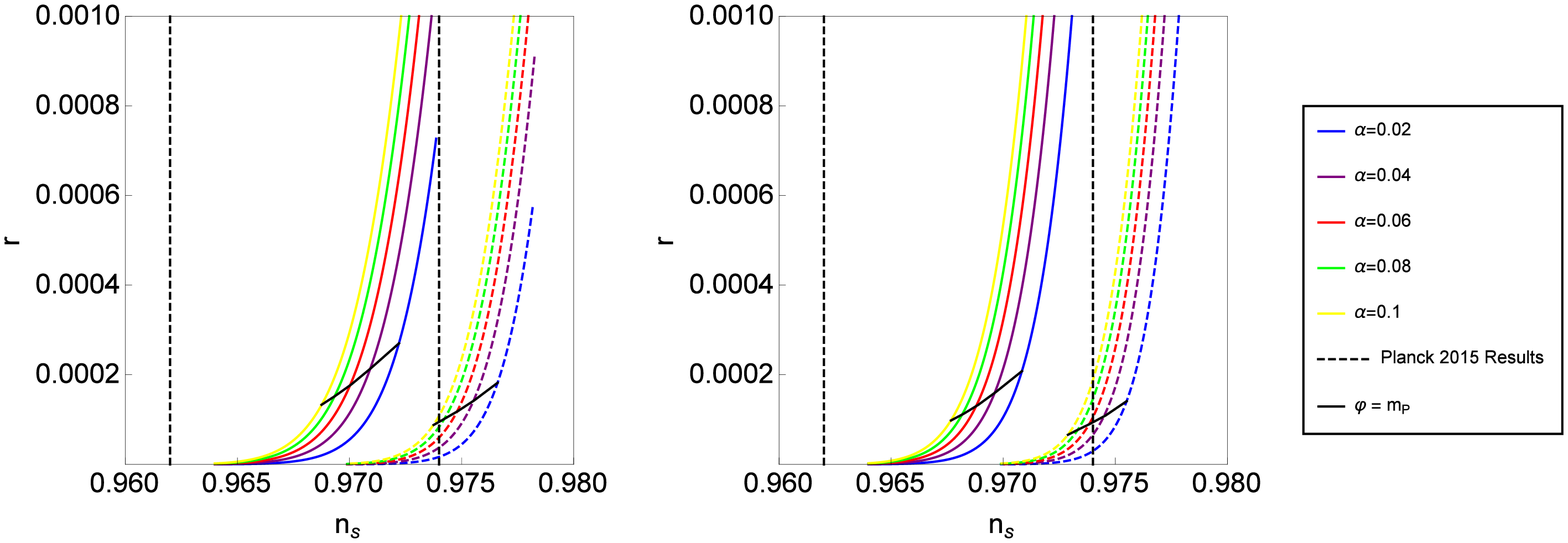}
\caption{The predictions of power-law plateau inflation in a graph depicting the tensor 
to scalar ratio $r$ in terms of the spectral index $n_s$.
Solid (dashed) lines correspond to $N=50$ $(N=60)$ respectively. On the left we 
have $q=1$ and on the right $q=4$. $n$ increases along the length of the lines,
counter-intuitively, \textit{right to left}. The vertical black dashed straight
lines represent the 1-$\sigma$ bounds on $n_s$ from the Planck data. The slanted
solid black line depicts the limit below which values of $\varphi$ are 
sub-Planckian.}
\label{fig:AllPlotsZoomed}
\vspace{-3cm}
\end{figure}

To test if our potential is a successful model we need to see if it produces 
satisfactory values for the tensor to scalar ratio and the spectral index when 
compared to the Planck observations~\cite{Ade:2015lrj}. We have four variables 
and parameters, namely $n$, $q$, $\alpha$, and $N$. To start with, we consider 
$N=50$ and $N=60$. We restrict $n$ and $q$ to integer values and keep $\alpha$ 
in the range 0.01 to 0.1 for our initial investigation. At this point we 
maintain $\varphi$ at sub-Planckian values; super-Planckian $\varphi$ will be 
considered later. We also ensure 
%\mbox{$\Big(\frac{\varphi}{m}\Big)^{n}>1$} at 
\mbox{$(\varphi/m)^{n}>1$} at 
all times. Fig.~\ref{fig:AllPlots} gives an overview of how $r$ and $n_{s}$ vary
for differing $n$, $q$ and $\alpha$, within this scope. 
Fig.~\ref{fig:AllPlotsZoomed} shows the relevant sections of the graph in more 
detail. The slanted black line marks the highest allowed $n$ value (non-integer)
for the relevant $q$, $\alpha$, $N$ combination. 


\vspace{5cm}

\begin{figure} [h]
    \begin{subfigure}[b]{0.5\linewidth}
    \centering
    \includegraphics[width=\linewidth]{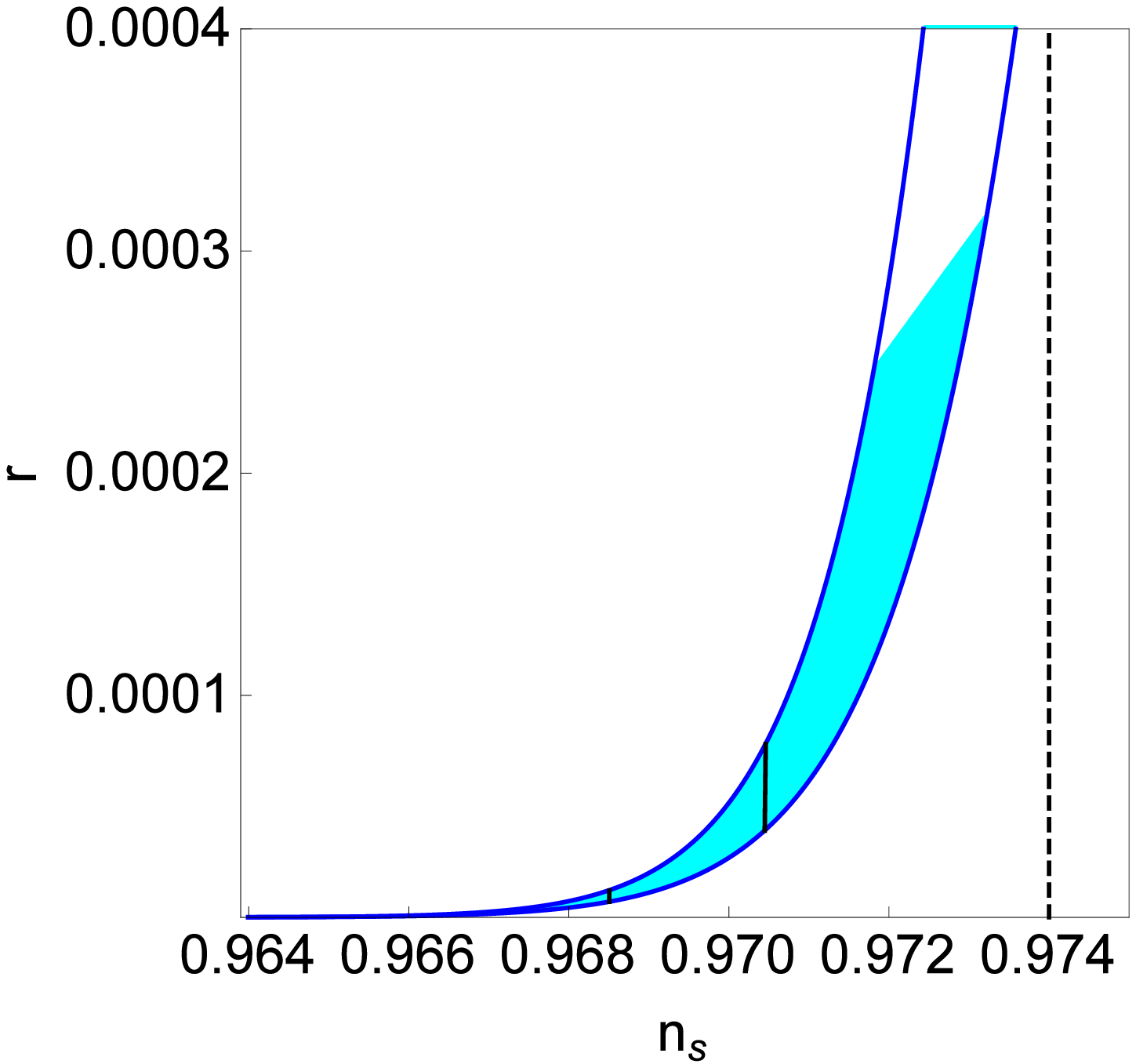} 
    \caption{$\alpha=0.01$} 
    \label{fig:WindowedAlphaZoom:a} 
    \vspace{4ex}
  \end{subfigure}%% 
  \begin{subfigure}[b]{0.5\linewidth}
    \centering
    \includegraphics[width=\linewidth]{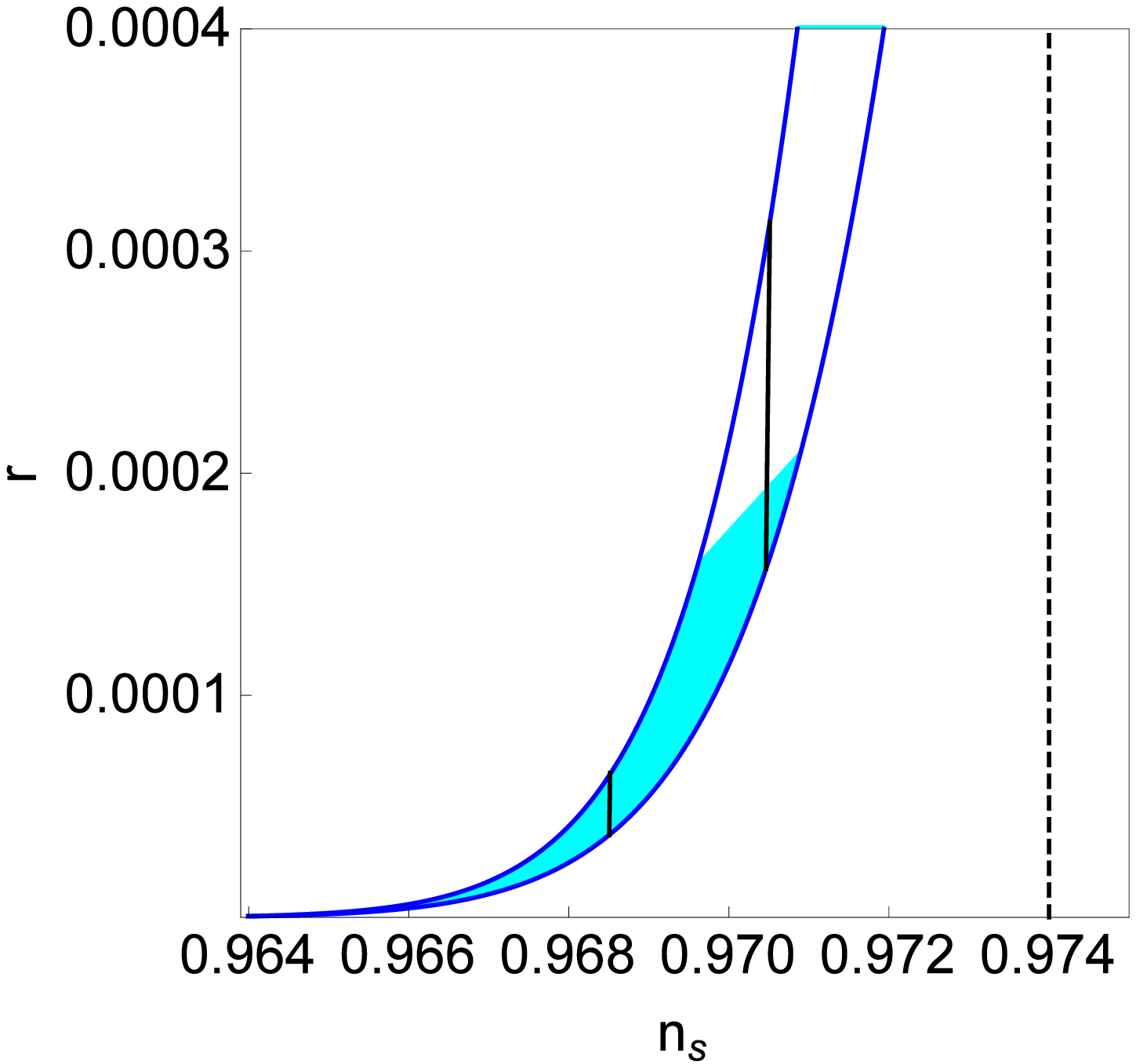} 
    \caption{$\alpha=0.04$} 
    \label{fig:WindowedAlphaZoom:b} 
    \vspace{4ex}
  \end{subfigure} 
  \begin{subfigure}[b]{0.5\linewidth}
    \centering
    \includegraphics[width=\linewidth]{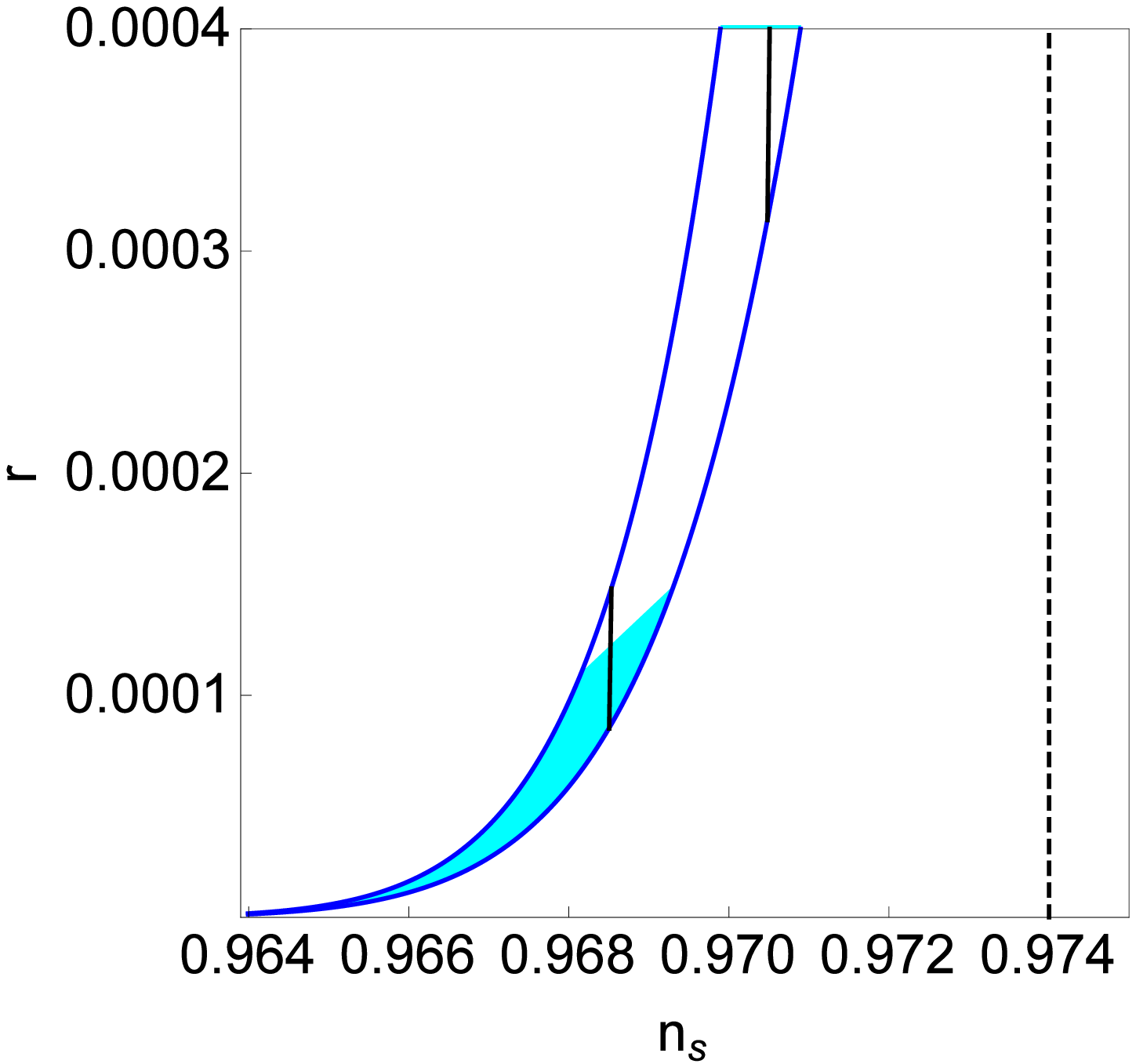} 
    \caption{$\alpha=0.08$} 
    \label{fig:WindowedAlphaZoom:c} 
  \end{subfigure}%%
  \begin{subfigure}[b]{0.5\linewidth}
    \centering
    \includegraphics[width=\linewidth]{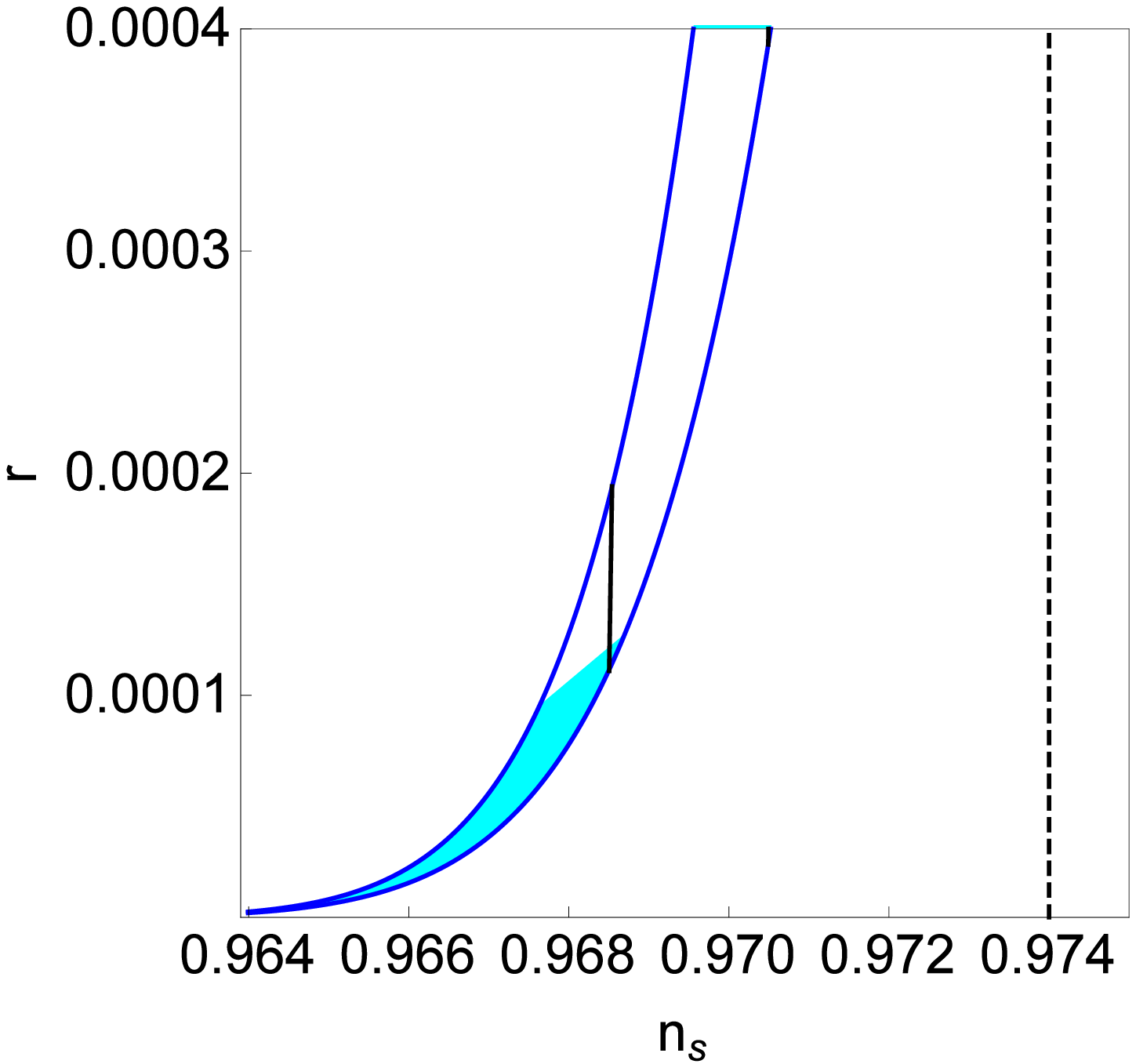} 
    \caption{$\alpha=0.1$} 
    \label{fig:WindowedAlphaZoom:d} 
  \end{subfigure} 
  \caption{The predictions of power-law plateau inflation in a graph depicting the tensor 
to scalar ratio, $r$, in terms of the spectral index $n_s$, assuming $N=50$. 
The blue shaded region in each plot shows the range of values for $r$ and 
$n_{s}$ for the acceptable choices of $n$ and $q$ which maintain sub-Planckian 
$\varphi$. The lower blue line in each window shows $q=1$ and the 
higher $q=4$, however $n$ varies along the length of each line; $n=1$ is the 
highest point of each line and $n=10$ the lowest. The vertical solid black 
lines represent $n=2$ and $n=3$. Note that, counter-intuitively, $n=2$ 
is the line on the right-hand side and $n=3$ on the left. Note also that, in 
panel~(d),
%Fig. \ref{fig:WindowedAlphaZoom:d} 
the $n=2$ line is off the top of the graph and only the $n=3$ line is visible. 
The vertical black dashed line is the upper $n_{s}$ Planck 1-$\sigma$ bound.}  
  \label{fig:WindowedAlphaZoom} 
\end{figure}

We can see from Fig.~\ref{fig:AllPlotsZoomed} that sensible $n_{s}$ values exist
in most cases but it would be better if $r$ were maximised to make contact with
future observations. From the graph, it is clear that increasing $n$ decreases 
$r$. However, lowering $n$ increases $\varphi$ so we need to balance this whilst
staying sub-Planckian. Increasing $q$ increases $r$ only marginally for all 
values of $n$ except $n=1$, but as shown in Fig.~\ref{fig:WindowedAlphaZoom}, 
$n=1$ is ruled out. However, again, this also increases $\varphi$. We can also 
see that a larger value of $\alpha$ is better for maximising our $r$ values, as
expected since $r \propto \alpha^{\frac{2n}{n+2}}$ to lowest order in 
Eq.~\eqref{eq:firstorderr}, this again also increases $\varphi$ though. 
Table~\ref{table:obsfromgraphs} summarises these results. 
%and they are visualised in Fig. \ref{fig:VaryingEffects}.

\begin{table}[h]
\begin{center}
\begin{tabular}{|c||c|c|c|}
\hline  & $n$ increased & $q$ increased & $\alpha$ increased \\ \hline
\hline $n_{s}$ & 
Substantially decreased & Marginally decreased & Complex pattern \\ 
\hline $r$ & 
Decreased (order of mag.) & Increased marginally (if $n\neq1$) & Increases \\ 
\hline $\varphi$ & 
Decreased & Increased, impact reduced as $n$ grows & Increases \\ 
\hline 
\end{tabular} 
\end{center}
\caption{Varying effects of the variables/parameters.}
\label{table:obsfromgraphs}
\end{table}

%\clearpage

We can see from the graphs that irrespective of the $n$, $q$ and $\alpha$ 
choices a lower value of $N$ always produces a higher value of $r$, brings 
$n_{s}$ more acceptably within the Planck bounds and actually decreases 
$\varphi$ too. The value of $N$ is discussed in the next section. For now we 
will focus on the $N=50$ results, implying a low reheating temperature.
Fig.~\ref{fig:WindowedAlphaZoom} combines the variations of $\alpha$, $n$ and 
$q$ with the effects on $\varphi$ to show how the allowed parameter space for 
$n_{s}$ and $r$ varies.
%Fig. \ref{fig:WindowedAlphaZoom} focuses on the areas of sub-Plankcian values of $\varphi$ and 
%
All values of $\alpha$ keep $n_{s}$ within the Planck bounds, the upper 
1-$\sigma$ bound of which is shown by the black dashed lines in the panels
of Fig.~\ref{fig:WindowedAlphaZoom}, while the lower bound is out of frame.  

At this stage, considering only integer values of $n$ and $q$, it seems that 
$n=2$ is the best choice, as lower $n$ values give higher $r$ results (when 
allowed by $\alpha$) and $n=1$ is ruled out for all the sub-Planckian cases we 
are currently considering. Table~\ref{table:maxalphaN50} tabulates the values 
of $\alpha$ and $q$ which maximise $r$ for each $n$ and validates the 
expectation that $n=2$ provides the highest value. 
%(possibly, combined with $q=2$ and $\alpha = 0.03$). 
Higher values of $q$ do not provide the highest result because their $\alpha$ 
values are capped by the \mbox{$\varphi<m_{P}$} criterion.

\begin{table}[h]
\begin{center}
\begin{tabular}{|c|c|c|c|c|}
\hline  $n$ & $q$ & Max allowed $\alpha$ & $n_{s}$ & $r$ \\ \hline 
\hline  2 & 1 & 0.04 & 0.970463 & 0.000157 \\ 
\hline  2 & 2 & 0.03 & 0.970474 & 0.000166 \\
\hline  2 & 3 & 0.02 & 0.970472 & 0.000136 \\
\hline  2 & 4 & 0.02 & 0.970478 & 0.000157 \\
\hline  3 & 1 & 0.10 & 0.968504 & 0.000111 \\
\hline  3 & 2 & 0.08 & 0.968518 & 0.000112 \\
\hline  3 & 3 & 0.07 & 0.968523 & 0.000113 \\
\hline  3 & 4 & 0.06 & 0.968524 & 0.000105 \\
\hline  4 & 1 & 0.16 & 0.967203 & 0.000080 \\
\hline  4 & 2 & 0.14 & 0.967218 & 0.000084 \\
\hline  4 & 3 & 0.12 & 0.967223 & 0.000079 \\
\hline  4 & 4 & 0.11 & 0.967225 & 0.000077 \\
\hline
\end{tabular} 
\end{center}
\caption{Maximum values that $\alpha$ can take %(to nearest $0.01$) 
for specific combinations of $n$ and $q$ when $N = 50$ whilst $\varphi$
remains sub-Planckian, and the corresponding values of $n_{s}$ and $r$.}
\label{table:maxalphaN50}
\end{table}

\clearpage
\section{\boldmath Lowering $N$}\leavevmode\\
So far we have examined the results for $n_{s}$ and $r$ when $N=50$ but perhaps
it is possible to lower the value of $N$ further if we can manipulate when the 
epoch of reheating began. This section will investigate this and then examine 
how a period of thermal inflation may affect $N$. 

\subsection{Low Reheating Temperature}
We start from the familiar equation:
%
%\begin{equation}
%2H_{k}^{-1} = H_{*}^{-1}\Big(\frac{a_{\mathrm{end}}}{a_{*}}\Big)\Big(\frac{a_{\mathrm{reh}}}{a_{\mathrm{end}}}\Big)\Big(\frac{a_{\mathrm{eq}}}{a_{\mathrm{reh}}}\Big)\Big(\frac{a_{k}}{a_{\mathrm{eq}}}\Big)\,,
%\end{equation}
%we find:
%
\begin{equation}
N_{*} = 62.8 + \mathrm{ln}\Big(\frac{k}{a_{0}H_{0}}\Big) + 
\mathrm{ln}\Big(\frac{V_{\mathrm{end}}^{1/4}}{10^{16}\mathrm{GeV}}\Big) + 
\frac{1}{3}\mathrm{ln}\Big(\frac{g_{*}}{106.75}\Big) -\Delta N\,,
\end{equation}
where $N_*$ is the number of remaining e-folds of inflation when the 
cosmological scales exit the horizon, \mbox{$k=0.002\,$Mpc$^{-1}$} is the pivot 
scale, $V_{\mathrm{end}}$ is the energy density at the end of inflation, $g_*$ 
is the number of effective relativistic degrees of freedom at reheating and
\begin{equation}
\Delta N \simeq 
\frac{1}{3}\mathrm{ln}\Big(\frac{V_{*}^{1/4}}{T_{\mathrm{reh}}}\Big)\,,
\end{equation}
where $V_{*}$ is the energy density when the cosmological scales exit the 
horizon and we assume that between the end of inflation and reheating the 
Universe is dominated by the inflaton condensate coherently oscillating in a 
quadratic potential around its vacuum expectation value. The latest data gives 
\mbox{$V_{*}^{1/4}<2\times 10^{16}\,\mathrm{GeV}$}~\cite{Ade:2015lrj}. Assuming 
that $T_{\mathrm{reh}}$ is greater than the electroweak (EW) scale
(to allow EW-baryogenesis), we have:
\begin{equation}
\Delta N \lesssim 
\frac{1}{3}\mathrm{ln}
\Big(\frac{V^{1/4}_{\rm *\;max}}{T_{\mathrm{reh}}^{\mathrm{min}}}\Big) \simeq 
\frac{1}{3}\mathrm{ln}\Big(
\frac{2\times 10^{16}\, \mathrm{GeV}}{200\, \mathrm{GeV}}\Big)
\simeq 10.8\,,
\end{equation}
which validates our choice to use the $N=50$ value, but does not 
introduce any lower values.

\subsection{Thermal Inflation}\leavevmode\\
Thermal inflation is a brief period of inflation possibly occurring after the 
reheating from primordial inflation. This second bout of inflation would allow 
the further reduction of the number of primordial e-folds of inflation. It 
occurs in the period between a false vacuum dominating over thermal energy 
density and the onset of a phase transition which cancels the false vacuum, 
when the thermal energy density falls to a critical value. The dynamics of
thermal inflation are determined by a so-called flaton field,  which is
typically a supersymmetric flat direction lifted by a negative soft 
mass~\cite{flaton}. The scalar potential for the thermal flaton field is of the
form:
\begin{equation}
V = V_{0} - \frac{1}{2}m^{2}\phi^{2} + \frac{1}{2}g^{2}T^{2}\phi^{2} + \cdots 
\label{eq:thermalinfl}
\end{equation}
where $m$ is the tachyonic mass of the field and $g$ is the coupling to the 
thermal bath, with the ellipsis denoting non-renormalisable terms that stabilise
the zero-temperature potential.\footnote{%
There is no self-interaction quartic term for flaton fields~\cite{flaton}.}
From the above, the effective mass-squared of the flaton field is
\begin{equation}
m_{\mathrm{eff}}^{2} = g^{2}T^{2} - m^{2}.
\label{eq:meffthermal}
\end{equation}
For high temperatures $m_{\rm eff}^2$ is positive and the flaton field is driven 
to zero, where $V=V_0>0$. This false vacuum density dominates when the thermal
bath temperature drops to the value $T_1$ such that
\begin{equation}
\rho_T(T_1)=\frac{\pi^{2}}{30}g_{*}T_{1}^{4}\equiv V_{0} 
\;\Rightarrow\; T_{1}=\Big(\frac{30}{\pi^{2}g_{*}}\Big)^{1/4}V_{0}^{1/4}
\sim V_{0}^{1/4},
\label{eq:thermaldensity}
\end{equation}
where $\rho_T$ is the density of the thermal bath. At $T_1$, thermal inflation 
begins and it continues until the temperature decreases enough that 
$m_{\rm eff}^2$ ceases to be positive. This critical temperature is
\begin{equation}
m_{\mathrm{eff}}^{2}(T_{2})\equiv 0 \; \Rightarrow \; T_{2} = m/g\,,
\end{equation}
when the effective mass-squared becomes tachyonic and a phase transition occurs,
which sends the flaton field to the true vacuum, thereby terminating thermal
inflation.

The total e-folds of thermal inflation are estimated as
\begin{equation}
N_{T} = \mathrm{ln}\Big(\frac{a_{\mathrm{end}}}{a_{\mathrm{beg}}}\Big) = 
\mathrm{ln}\Big(\frac{T_{1}}{T_{2}}\Big) = 
\frac{1}{4}\mathrm{ln}\Big(\frac{30}{\pi^{2}g_{*}}\Big) + 
\mathrm{ln}\frac{gV_{0}^{1/4}}{m}\,,
\label{eq:thermal1}
\end{equation}
where $a_{\mathrm{beg}}$ ($a_{\mathrm{end}}$) is the scale factor of the Universe at 
the beginning (end) of thermal inflation.

We also know
\begin{equation}
g\leq1, \qquad m^{2}\langle\phi\rangle^2 \sim V_{0}, \qquad 
\langle\phi\rangle \leq m_{P}
\end{equation}
resulting in 
\begin{equation}
V_{0}^{\mathrm{max}} \sim (m\, m_{P})^{2}
\end{equation}
Therefore:
\begin{equation}
N_T\leq
\underbrace{\frac{1}{4}\mathrm{ln}\Big(\frac{30}{\pi^{2}g_{*}}\Big)}_{\simeq 0}
+\mathrm{ln}\Big(g\sqrt{\frac{m_{P}}{m}}\Big) 
\leq \frac{1}{2}\mathrm{ln}\Big(\frac{m_{P}}{m}\Big)\,.
\end{equation} 
The maximum value we can allocate to $N_{T}$ and hence reduce our e-folds of primordial inflation by (since $N_*\rightarrow N_*-N_T$), would arise from 
minimising $m$. The minimum $m$ is given by the electroweak scale 
\mbox{$m_{_{\rm EW}}\sim 1\,\mathrm{TeV}$} since $\phi$ particles are not observed
in the LHC. Using these values we obtain:
\begin{equation}
N_{T} \leq \frac{1}{2}\mathrm{ln}\frac{m_{P}}{m_{_{\mathrm{EW}}}} \simeq 17
\label{eq:thermalis17}
\end{equation}

So, considering that 17 e-folds of thermal inflation occurred, we may lower the
value of $N_*$ down to $33$ or so. However, this will affect the combinations of
$n$, $q$ and $\alpha$ that are still able to maintain sub-Planckian $\varphi$.
%Fig. \ref{fig:difNintegern} highlights the options for integer values of $q$ and $\alpha$ values in steps of $0.01$ over the range $N=33$ to the maximum allowed $N$ value within the specified limits. 
From the previous graphs we know that $n=2$ gives us higher values for $r$ than
any higher integer $n$ value (since $n=1$ 
%gives substantially higher values but is disallowed for all combinations of variables except one because it 
violates the $\varphi<m_P$ bound). 
%It is acceptable for $n=1$, $q=1$, $\alpha=0.01$, $N=33$ but its result for $n_{s}$ ($0.961$) is outside of the Planck results so it has been omitted.
For %all of the series with 
$n=2$, 
we find $39 \leqslant  N_* \leqslant 56$ are within the range of values in the 
Planck 2015 data for $n_{s}$. 
%For every series lower values of $N$ produce higher values of $r$, as before. Based on these graphs the best combination of variables (using integer values for $N$, $n$ and $q$) is $N=39$, $n=2$, $q=3$, $\alpha=0.03$. 
We also find that increasing $q$ has minimal effect on $r$. Hence, for 
simplicity, we set $q=1$.
%However, Fig.\ref{fig:difNintegern} highlights how minimal an effect an increase of $q$ has on $r$. For aesthetic reasons a $q$ value of $1$ is preferred. 
Because of this we highlight $n=2$ and $q=1$ (with $\alpha=0.05$) 
as our best result, corresponding to the potential:
\begin{equation}
V = V_{0}\,%\Big(
\frac{\varphi^{2}}{\varphi^{2} + m^{2}}
%\Big)
\,.
\end{equation}
The $n_{s}$ and $r$ results for this combination of variables are shown in 
Table~\ref{tab:running2}. In this table,
%
%This means the series $n=2$, $q=3$, $\alpha=0.03$ is the most promising for maximising $r$ as long as the calculated $N$ value falls within its allowed range which is $39 \leqslant  N \leqslant 45$. If this is not the case, the next best result has a larger $N$ range, namely $39 \leqslant  N \leqslant 49$ which is for the series $n=2$, $q=1$, $\alpha=0.05$. 
%The first series to allow any value of $N$ that puts $n_{s}$ into the Planck range is $n=2$, $q=2$, $\alpha=0.03$.
%
%\begin{figure}
%\centering
%\includegraphics[width=\textwidth]{difNintegerfinal.eps}
%\caption[]{$N$ increases left to right, always starting at 33 but max value varies, $n=2$ unless explicitly stated, all crosses shown are within the limit of $\varphi < m_{p}$ {hence why some series do not include as many $N$ values}}
%\label{fig:difNintegern}
%\end{figure}
%
we also include results for the running of the spectral index which the Planck 
observations found as 
\mbox{$\frac{\mathrm{d}n_{s}}{\mathrm{dln}k}=-0.003 \pm 0.007$}~%
\cite{Ade:2015lrj}. 
%We can calculate the possible values in our case using the equation: 
%$\frac{\mathrm{d}n_{s}}{\mathrm{dln}k} = -\frac{\mathrm{d}n_{s}}{\mathrm{d}N}$:
%
%\begin{equation}
%\frac{\mathrm{d}n_{s}}{\mathrm{dln}k} = -\frac{\mathrm{d}}{\mathrm{d}N}\Big[1 - 2\frac{(n+1)}{(n+2)}\Big(N + \frac{n+1}{n+2}\Big)^{-1}\Big\{1 + \frac{(n^{2}-2nq-n)}{2(n+1)} \Big[\frac{n(n+2)q}{\alpha^{2}} \Big(N + \frac{n+1}{n+2}\Big)\Big]^{\frac{-n}{n+2}}\Big\}\Big]
%\end{equation}
In our model,
%\begin{align}
%\frac{\mathrm{d}n_{s}}{\mathrm{dln}k} = 
%-\frac{2(n+1)}{(n+2)}\Big(N + \frac{n+1}{n+2}\Big)^{-2} 
%\Big\{1+\frac{(n^{2}-2nq-n)}{2(n+1)}
%\Big(\frac{n(n+2)q}{\alpha^{2}}\Big)^{-\frac{n}{n+2}} \nonumber \\ 
%\times \quad \Big[\frac{n}{n+2}\Big(N + \frac{n+1}{n+2}\Big)^{-\frac{n}{n+2}} + 
%\Big(N +  \frac{n+1}{n+2}\Big)^{\frac{2}{n+2}}\Big]\Big\}\,.
%\end{align}
%Results are presented
\begin{equation}
\frac{{\rm d}n_s}{{\rm d}\ln k}=
-2\frac{n+1}{n+2}\left(N+\frac{n+1}{n+2}\right)^{-2}
\left\{1+\frac{n^2-2nq-n}{n+2}
\left[\frac{n(n+2)q}{\alpha^2}\left(N+\frac{n+1}{n+2}\right)
\right]^{-\frac{n}{n+2}}
\right\}
\end{equation}

As shown in Table \ref{tab:running2}, our model is in excellent agreement 
with the Planck observations.

\begin{table}[h]
\begin{center}
\begin{tabular}{|c|c|c|c|c|c|c|}
\hline $n$ &$q$&$\alpha$&$N$ & $n_{s}$  &$r$   & $\frac{\mathrm{d}n_{s}}{\mathrm{d}N}$\\ \hline
\hline 2   & 1 & 0.05   & 39 & 0.962299 & 0.000283 & -0.00095   \\
\hline 2   & 1 & 0.05   & 49 & 0.969874 & 0.000202 & -0.00061   \\
\hline 
\end{tabular} 
\end{center}
\caption{Results for $n_{s}$, $r$ and the running of the spectral index for the
case $n=2$, $q=1$, $\alpha=0.05$ showing the extremal allowed values of $N$ for
completeness ($39 \leqslant  N_* \leqslant 49$).}
\label{tab:running2}
\end{table}

\section{Single Mass Scale}\leavevmode\\
In an attempt to make the model more economic, we consider that our model is characterised by a single mass 
scale $M$, such that $V_0^{1/4}=m\equiv M$. Under this assumption, the model is 
more constrained and more predictive. The scalar potential is
\begin{equation}
V = M^{4}\Big(\frac{\varphi^{n}}{\varphi^{n} + M^{n}}\Big)^{q}\,.
\label{eq:potentialMm}
\end{equation}
The inflationary scale is determined by the COBE constraint:
\begin{equation}
\sqrt{\mathscr{P}_{\zeta}} = \frac{1}{2\sqrt{3}\pi}\frac{V^{\frac{3}{2}}}{m_{P}^{3}V'}\,,
\label{eq:inflscale}
\end{equation}
where $\mathscr{P}_{\zeta} = (2.208 \pm 0.075) \times 10^{-9}$. 
is the spectrum of the scalar curvature perturbation. 
From Eqs.~\eqref{eq:main} and \eqref{eq:firstorderphioverm} we obtain
\begin{equation}
\Big(\frac{M}{m_{P}}\Big)=
(2\sqrt{3}\pi nq\sqrt{\mathscr{P}_{\zeta}})^{\frac{n+2}{n+4}}
\Big[n(n+2)q\Big(N+\frac{n+1}{n+2}\Big)\Big]^{-\frac{n+1}{n+4}}.
\end{equation}

%\begin{table}[h]
%\begin{center}
%\begin{tabular}{|c|c|c|c|c|}
%\hline $n$   & $q$ & $N$  & $M\,(\mathrm{GeV})$                        & $\alpha$ \\
%\hline 2   & 1 & 39 &  $9.45\mathrm{x10}^{15}$ & 0.0039     \\
%\hline 2   & 1 & 49 &  $9.10\mathrm{x10}^{15}$ & 0.0037     \\
%\hline 
%\end{tabular} 
%\end{center}
%\caption{Calculating $M$ and subsequently $\alpha$}
%\label{tab:Mm}
%\end{table}

Inputting the calculated values of $\alpha=M/m_P$ for when 
$M=m=V_{0}^{1/4}$ into the equations for $n_{s}$, $r$ and $\frac{dn_s}{d\ln k}$,
we find the results shown in Table \ref{tab:Results when Mm}. 

%Unfortunately, in this case, streamlining the potential by only invoking one mass scale is not beneficial to our results for $n_{s}$ and $r$.

 \begin{table}[h]
 \begin{center}
 \begin{tabular}{|c|c|c|c|c|c|c|}
 \hline $n$ &$q$&$N$ & $n_{s}$ & $r$ %($\times 10^{-6}$) 
& $\frac{\mathrm{d}n_{s}}{\mathrm{d}\,\mathrm{ln}k}$ & 
$M$ ($\times 10^{15}\,$GeV)\\ \hline
 \hline 2   & 1 & 39 & 0.962267& 3.2$\times 10^{-6}$  & -0.00095 &
1.39$\pm$0.01 \\
 \hline 2   & 1 & 49 & 0.969851& 2.1$\times 10^{-6}$  & -0.00061 &
1.24$\pm$0.01 \\
  \hline 
 \end{tabular} 
 \end{center}
 \caption{Values of $n_{s}$, $r$, $\frac{dn_s}{d\ln k}$ and $M$ when
$m=V_{0}^{1/4}\equiv M$.}
 \label{tab:Results when Mm}
 \end{table}
Whilst $n_{s}$ is well within the Planck
bounds, the values for $r$ have dropped considerably due to the reduction in 
\mbox{$\alpha=M/m_P\ll 1$}, which is expected as 
\mbox{$r \propto \alpha^{\frac{2n}{n+2}}$}.
Note that we find \mbox{$M\sim 10^{15}\,$GeV}, which is close to
the scale of grand unification as expected.

\section{Large-field Power-law Plateau Inflation}
\begin{figure}[h]
	\centering
	\includegraphics[width=\textwidth]{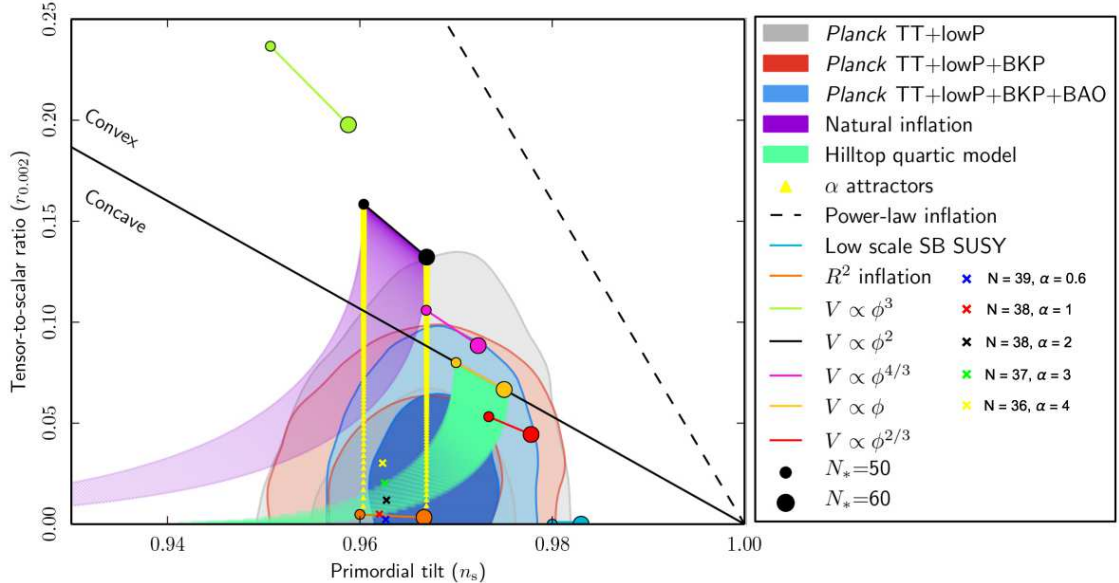}
	\caption{Super-Planckian power-law plateau inflation values of $n_{s}$ and $r$ 
		superimposed on the Planck Graph.} 
	\label{fig:OurResultsonPlanck}
\end{figure}

\begin{figure}[h]
	\centering
	\includegraphics[width=\textwidth]{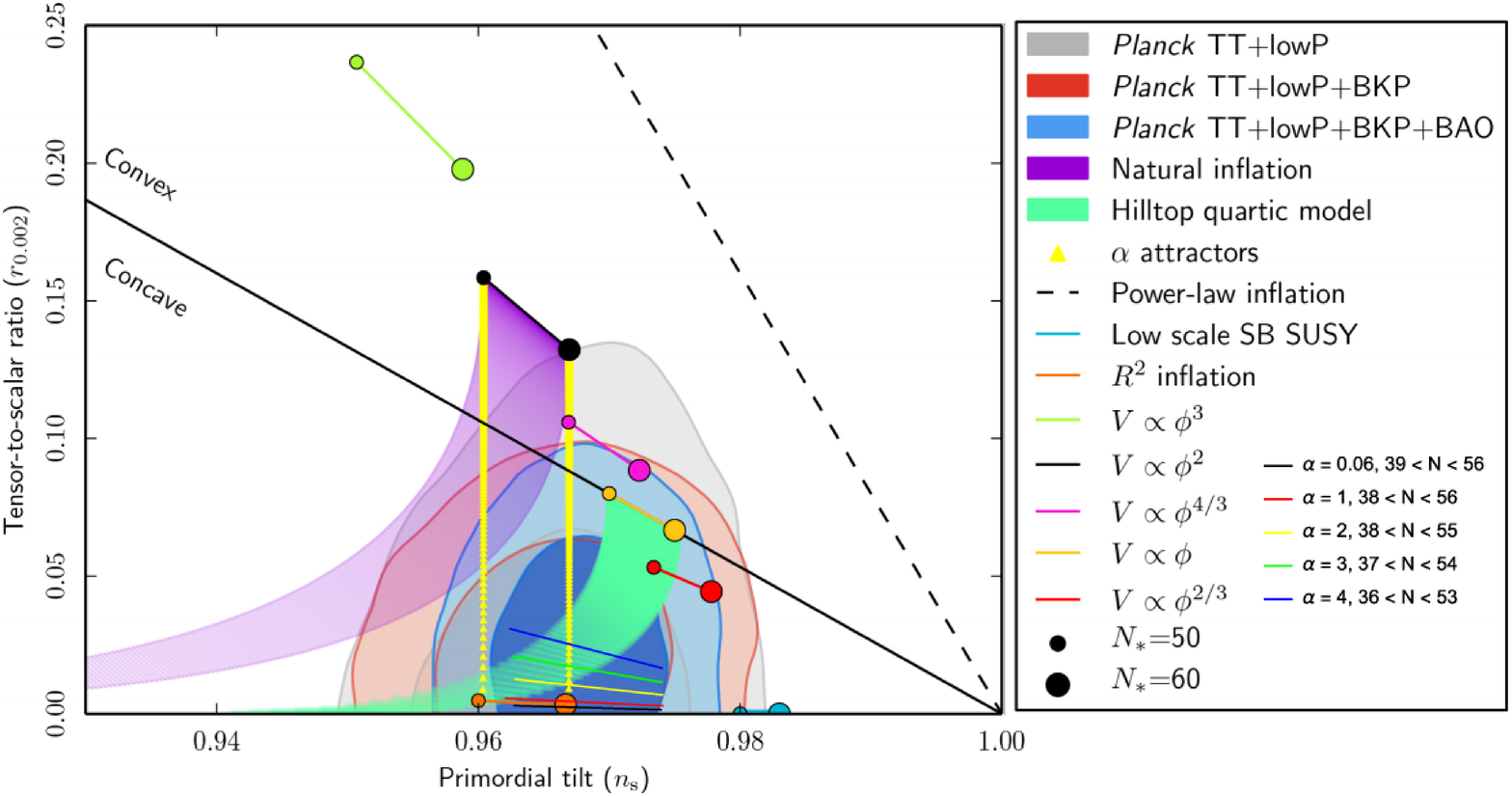}
	\caption{Super-Planckian power-law plateau inflation values of $n_{s}$ and $r$ 
		superimposed on the Planck Graph, with $N$ varying between its min and max 
		values for which $n_{s}$ is within the Planck 1-$\sigma$ bounds.} 
	\label{fig:OurResultsonPlanck2}
	%\vspace{-5cm}
\end{figure}

\leavevmode\\
In this section, we consider $\alpha$ no longer capped by ensuring 
$\varphi<m_{P}$ but we still satisfy the bound \mbox{$(\varphi/m)^n>1$}. 
Higher $\alpha$ values produce higher $r$ values but also higher $n_{s}$ values
so we must be wary our $n_{s}$ results do not migrate outside of the Planck 
bounds. To mitigate this, a lower $N$ value is again better, so we consider 
values down to 33. Table~\ref{tab:Maximising r} presents the best values of 
$n_{s}$ and $r$ for a combination of $N$ and $\alpha$ values. To demonstrate how
our model improves the tensor to scalar ratio and spectral index without any 
fine tuning\footnote{This is because both mass scales assume natural values; 
\mbox{$m\simeq m_P$} and $V_0^{1/4}$ is of the scale of grand unification.}, Table~\ref{tab:final super values for N 50 and 60} shows the 
results for $N=50$ and 60 and how they sit inside the Planck bounds. 
Fig.~\ref{fig:OurResultsonPlanck} shows the best results for five $\alpha$ 
values with their respective best $N$ values. Fig.~\ref{fig:OurResultsonPlanck2}
shows the range of our results for the same $\alpha$ values over all allowed 
$N$ values which maintain $n_{s}$ within the 1-$\sigma$ Planck bounds, fitting 
perfectly into the Planck parameter space. If we allow our $n_{s}$ results to 
extend into the 2-$\sigma$ Planck parameter space we can also incorporate 
$N=60$.

\begin{table}[h]
\begin{center}
\begin{tabular}{|c|c|c|c|c|c|}
\hline $\alpha$ & $N$ & $n_{s}$  & $r$      & $\frac{\varphi}{m}$  &  
$\frac{\mathrm{d}n_{s}}{\mathrm{d}\,\mathrm{ln}k}$       \\ \hline
\hline 0.6      & 39  & 0.962687 & 0.003500 & 5.36                 & -0.00093  \\ 
%\hline 0.7      & 39  & 0.962758 & 0.004105 & 4.95                 & -0.00045  \\ 
\hline 0.8      & 39  & 0.962828 & 0.004717 & 4.62                 & -0.00093  \\ 
%\hline 0.9      & 39  & 0.962899 & 0.005335 & 4.34                 & -0.00031  \\ 
\hline 1        & 38  & 0.962023 & 0.006169 & 4.08                 & -0.00097 \\ 
\hline 2        & 38  & 0.962756 & 0.013058 & 2.80                 & -0.00094 \\ 
\hline 3        & 37  & 0.962551 & 0.021450 & 2.20                 & -0.00096 \\ 
\hline 4        & 36  & 0.962358 & 0.031315 & 1.83                 & -0.00098 \\ 
%\hline 5        & 35  & 0.962177 & 0.042861 & 1.57                 & 0.00302 \\ 
\hline 6        & 34  & 0.962012 & 0.056328 & 1.37                 & -0.00102 \\ 
%\hline 7        & 34  & 0.962875 & 0.068615 & 1.22                 & 0.00489 \\ 
\hline 8        & 33  & 0.962768 & 0.085796 & 1.08                 & -0.00100 \\ 
\hline 
\end{tabular} 
\end{center}
\caption{Values of $n_{s}$ and $r$, with super-Planckian values of $\varphi$ 
(ensuring $\varphi>m$), $n=2$ and $q=1$. $N$ values chosen to 
maximise $r$ whilst keeping $n_{s}$ within the 1-$\sigma$ Planck bound.
Note, as well, that because Planck observations suggest 
\mbox{
$-0.010\leqslant\frac{\mathrm{d}n_{s}}{\mathrm{d ln}k}\leqslant 0.004$}
at 1-$\sigma$ \cite{Ade:2015lrj}, the running of the spectral index also matches observations.}
\label{tab:Maximising r}
\end{table}

\begin{table}
\begin{center}
\begin{tabular}{|c|c|c|c|c|}
\hline $N$ & $\alpha$ & $n_{s}$  & $r$ & 
$\frac{\mathrm{d}n_{s}}{\mathrm{d ln}k}$ \\ \hline
\hline 50  & 1        & 0.970932 & 0.004106 & -0.00057 \\ 
\hline 50  & 2        & 0.971421 & 0.008600 & -0.00055 \\ 
\hline 50  & 3        & 0.971910 & 0.013482 & -0.00054 \\ 
\hline 50  & 4        & 0.972399 & 0.018753 & -0.00052 \\ 
\hline 50  & 5        & 0.972888 & 0.024412 & -0.00051 \\ \hline
\hline 60  & 1        & 0.975682 & 0.003122 & -0.00040 \\ 
\hline 60  & 2        & 0.976055 & 0.006515 & -0.00039 \\ 
\hline 60  & 3        & 0.976429 & 0.010180 & -0.00038 \\ 
\hline 60  & 4        & 0.976802 & 0.014115 & -0.00037 \\ 
\hline 60  & 5        & 0.977175 & 0.018321 & -0.00036 \\ 
\hline 
\end{tabular}
\end{center}
\caption{Values of $n_{s}$ and $r$, with super-Planckian values of $\varphi$ 
(ensuring $\varphi>m$) for $N=50$ and $N=60$.}
\label{tab:final super values for N 50 and 60}
\end{table} 

\pagebreak


\pagebreak

\section{Supergravity Toy Model}
In this section we will present a toy-model in supergravity (SUGRA) which can
produce the scalar potential of Plateau Inflation for $n=2$ and $q=1$\footnote{This is also a model of Shaft Inflation \cite{shaft}.}. We will follow
the approach of Ref.~\cite{Sdual} but only in form, not assuming the same 
theoretical framework (hence, we only consider a toy model) and with an 
important difference: we consider a minimal K\"{a}hler potential so that we 
can avoid producing too large $n_s$, but retain the successes of the 
$n=2$, $q=1$ Plateau Inflation model.

\subsection{Global Supersymmetry}
At first, we consider only global supersymmetry (SUSY) and sub-Planckian fields.
We introduce the non-renormalisable superpotential:
\begin{equation}
W=\frac{S^2(\Phi_1^2-\Phi_2^2)}{2m}\,,
\label{W}
\end{equation}
where $S,\Phi_1,\Phi_2$ are chiral superfields and $m$ is a large, but 
sub-Planckian scale. The F-term scalar potential is then
\begin{equation}
V_F=\frac{|S|^2}{m^2}
\Big[|\Phi_1^2-\Phi_2^2|^2+|S|^2(|\Phi_1|^2+|\Phi_2|^2)\Big].
\label{VF0susy}
\end{equation}
The above potential is minimised when \mbox{$\Phi_1=\Phi_2$}. Rotating the 
fields in configuration space (assuming a suitable R-symmetry) we can introduce
a canonically normalised, real scalar field $\varphi$ such that 
\mbox{$|\Phi_1|=|\Phi_2|\equiv\frac12\varphi$}.
Then the scalar potential becomes
\begin{equation}
V_F=\frac{|S|^4\varphi^2}{2m^2}\,.
\label{VFsusy}
\end{equation}
We consider that there is also a D-term contribution to the scalar potential.
Mirroring Ref.~\cite{Sdual}, we take
\begin{equation}
V_D=\frac12(|S|^2-\sqrt 2M^2)^2,
\label{VD}
\end{equation}
where $M$ is the scale of a grand unified theory (GUT).
Thus, in total, the scalar potential reads
\begin{equation}
V=\frac{|S|^4\varphi^2}{2m^2}+\frac12(|S|^2-\sqrt 2M^2)^2.
\label{V0susy}
\end{equation}
Minimising the potential in the $S$ direction requires
\begin{equation}
\frac{\partial V}{\partial|S|}=0\;\Rightarrow\;
\langle|S|^2\rangle=\frac{\sqrt 2M^2}{1+\varphi^2/m^2}\,.
\end{equation}
Inserting the above in Eq.~\eqref{V0susy} we obtain
\begin{equation}
V=\frac{M^4\varphi^2}{m^2+\varphi^2}\,,
\label{Vsusy}
\end{equation}
which is the $n=2$, $q=1$ power-law plateau inflation model.

\subsection{Local Supersymmetry}
To confine ourselves in global SUSY we have to require that 
\mbox{$M<m<\varphi<m_P$}. Given that \mbox{$\log(m_P/M)\simeq 2$}, the available
parameter space is not a lot. However, generalising the above into SUGRA may 
allow super-Planckian values for the inflaton, while $m\simeq m_P$. 
In SUGRA, we continue to consider the superpotential in Eq.\eqref{W} and we
will also consider a minimal K\"{a}hler potential
\begin{equation}
K=|\Phi_1|^2+|\Phi_2|^2+|S|^2.
\end{equation}
Then the F-term scalar potential is
\begin{eqnarray}
V_F & = & \exp\left(\frac{|\Phi_1|^2+|\Phi_2|^2+|S|^2}{m_P^2}\right)\times
\nonumber\\
& & %\mbox{\hspace{-2cm}}
\left[\frac{|S|^2|\Phi_1^2-\Phi_2^2|^2}{m^2}
\left(1+\frac{2|S|^2}{m_P^2}+\frac{|S|^4}{4m_P^4}\right)+\right.\nonumber\\
& & \left.\frac{|S|^4(|\Phi_1|^2+|\Phi_2|^2)}{m^2}
\left(1+\frac{|\Phi_1^2-\Phi_2^2|^2}{4m_P^4}\right)
-3\,\frac{|S|^4|\Phi_1^2-\Phi_2^2|^2}{4m^2m_P^2}\right].
\end{eqnarray}
Considering that $|S|$ is sub-Planckian, since 
\mbox{$\langle|S|^2\rangle<\sqrt 2M^2$}, we have
\begin{equation}
V_F\simeq\exp\left(\frac{|\Phi_1|^2+|\Phi_2|^2}{m_P^2}\right)
\left[\frac{|S|^2|\Phi_1^2-\Phi_2^2|^2}{m^2}+
\frac{|S|^4(|\Phi_1|^2+|\Phi_2|^2)}{m^2}
\left(1+\frac{|\Phi_1^2-\Phi_2^2|^2}{4m_P^4}\right)\right].
\end{equation}
Again, the potential is minimised when \mbox{$\Phi_1=\Phi_2$}. Writing
\mbox{$|\Phi_1|=|\Phi_2|\equiv\frac12\varphi$}, we obtain
\begin{equation}
V_F=e^{^{\frac12(\varphi\!/\!m_{_{\!P}}\!)^2}}\frac{|S|^4\varphi^2}{2m^2}\,.
\label{VFsugra}
\end{equation}
We consider the same D-term contribution to the scalar potential, given in 
Eq.~\eqref{VD}. The total scalar potential is now
\begin{equation}
V=e^{^{\frac12(\varphi\!/\!m_{_{\!P}}\!)^2}}
\frac{|S|^4\varphi^2}{2m^2}+\frac12(|S|^2-\sqrt 2M^2)^2.
\label{V0sugra}
\end{equation}
Minimising the above along the $S$ direction, we find
\begin{equation}
\langle|S|^2\rangle=
\frac{\sqrt 2M^2}{1+e^{^{\frac12(\varphi\!/\!m_{_{\!P}}\!)^2}}(\varphi/m)^2}\,.
\end{equation}
Inserting this into Eq.~\eqref{V0sugra} we obtain
\begin{equation}
V=\frac{M^4\varphi^2}{e^{^{-\frac12(\varphi\!/\!m_{_{\!P}}\!)^2}}m^2+\varphi^2}\,.
\label{Vsugra}
\end{equation}
Our finding is technically different from Eq.~\eqref{Vsusy} because of the 
exponential factor in the first term in the denominator. However, this term is
important only when \mbox{$\varphi<m\simeq m_P$}, when the exponential is unity.
In the opposite case, when \mbox{$\varphi>m\simeq m_P$}, the first term in the 
denominator is negligible. So the exponential factor makes no difference and
the potential is practically the same with the one in Eq.~\eqref{Vsusy}. This 
is certainly so if the inflaton remains sub-Planckian (i.e. when $m<m_P$).

Using the potential in Eq.~\eqref{Vsugra} it can be checked that the 
$\eta$-problem of SUGRA inflation is overcome due to the D-term.
In fact, one can show that \mbox{$|\eta|\ll 1$}
%\begin{equation}
%\eta\simeq
%-\left(\frac{m}{m_P}\right)^2e^{^{-(\varphi\!/\!m_{_{\!P}}\!)^2}}\ll 1\,,
%\end{equation}
when \mbox{$\varphi>m_P\simeq m$}. However, for a super-Planckian inflaton,
things are different from power-law plateau inflation, since the SUGRA correction 
dominates. So, our SUGRA toy model is 
similar but not identical to the the $n=2$, $q=1$ power-law plateau inflation model.
%The difference can be seen in Fig.~?? where the predictions of our toy model 
%and the $n=2$, $q=1$ Plateau Inflation are shown for various choices of 
%$\alpha$.

To have an idea of the value of the observables in this toy model, we can
investigate the slow-roll parameters, which are found to be
\begin{eqnarray}
\epsilon & = & \frac{2}{x}\frac{\alpha^4}{(\alpha^2+e^{\frac12 x}x)^2}
\left(1+\frac12x\right)^2\quad{\rm and}\quad\\
\eta & = & \frac{1}{x}\frac{2\alpha^2}{\alpha^2+e^{\frac12 x}x}
\left[\frac{\alpha^2(1+2x)-3x}{\alpha^2+e^{\frac12 x}x}\left(1+\frac12 x\right)-
\frac12 x^2\right],
\end{eqnarray}
where \mbox{$x\equiv(\varphi/m_P)^2$} and $\alpha$ is given in 
Eq.~\eqref{alpha}. Then the spectral index is
\begin{equation}
n_s-1=2\eta-6\epsilon=\frac{4}{x}\frac{\alpha^4}{(\alpha^2+e^{\frac12 x}x)^2}
\left[\frac{\alpha^2(\frac12x-2)-3x}{\alpha^2+e^{\frac12 x}x}
\left(1+\frac12 x\right)-\frac12 x^2\right].
\end{equation}
Taking $\varphi>m_P\Rightarrow x\gg 1$ and $\alpha\sim 1$
simplifies the above considerably as
\begin{equation}
\epsilon\simeq\frac{\alpha^4}{2xe^x}
\quad{\rm and}\quad
\eta\simeq-\alpha^2 e^{-\frac12x},
\end{equation}
and it is evident that \mbox{$|\eta|\ll 1$} as already mentioned. In this limit,
it is easy to find
\begin{equation}
N=\frac14\int_{x_{\rm end}}^x\frac{\alpha^2+e^{\frac12x}x}{\alpha^2(1+\frac12x)}\,dx
\;\Rightarrow\;\alpha^2 N\simeq\exp(x/2)\,,
\end{equation}
where we have taken \mbox{$x\gg x_{\rm end}$}. Then, the observables become
\begin{equation}
n_s\simeq 1+2\eta\simeq 1-\frac{2}{N}\quad{\rm and}\quad
r=16\epsilon\simeq\frac{4}{\ln(\alpha^2N)N^2}
\end{equation}
These should be contrasted with the predictions of power-law plateau inflation given by
Eqs.~\eqref{eq:firstorderr} and \eqref{eq:firstorderns}. At lowest order and 
taking \mbox{$N\gg 1$} we find
\begin{equation}
n_s\simeq 1-\frac{3}{2N}\quad{\rm and}\quad
r\simeq\frac{\sqrt 2\alpha}{N^{3/2}}\,.
\end{equation}
We see that the predictions of our SUGRA toy-model are more pronounced with respect to $N$, with 
both the spectral index and the tensor to scalar ratio smaller. Also, the 
dependence of $r$ on $\alpha$ is more prominent in the case of power-law plateau 
inflation.
For more realistic values of the inflaton, however, where 
\mbox{$\varphi\sim m_P\sim m$} (i.e. \mbox{$x\sim 1$}) we would expect $n_s$ and
$r$ to lie inbetween the above extremes. Note that the predicted values are well
in agreement with the Planck data, in all cases.

\section{Conclusions}
We have studied in detail a new family of inflationary models called power-law plateau 
inflation. The models feature an inflationary plateau, which is approached 
in a power-law manner, in contrast to the popular Starobinsky/Higgs inflation
models (and their variants) but similarly to Shaft Inflation. We have shown that
power-law plateau inflation is in excellent agreement with Planck observations. 

To avoid supergravity corrections we mostly considered a sub-Planckian excursion
for the inflaton in field space. As expected, this resulted in very small values
for the ratio of the spectra of tensor to scalar curvature perturbation $r$.
In an attempt to improve our results and produce observable $r$ we have 
considered minimising the remaining number of e-folds of primordial inflation 
when the cosmological scales exit the horizon. To this end, we assumed late
reheating as well as a subsequent period of thermal inflation, driven by a 
suitable flaton field. We have managed to achieve
\mbox{$r\simeq 3\times 10^{-4}$} which might be observable in the future 
(see Table~\ref{tab:running2}).

For economy we have also investigated the possibility that our model is 
characterised by a single mass scale. We have found that the spectral index of 
the scalar curvature perturbation $n_s$ satisfies well the Planck 
observations but the model produces unobservable $r$.

Abandoning sub-Planckian requirements allows the model to achieve much larger
values of $r$. Indeed, for natural values of the mass scales (Planck and GUT 
scale), i.e. without fine-tuning, we easily obtain $r$ as large as a few 
percent (up to 9\%, see Table~\ref{tab:Maximising r}), which is testable in 
the near future. Our predicted values for $r$ and $n_s$ fall
comfortably within the 1-$\sigma$ bounds of the Planck observations, while
different models of the power-law plateau inflation family are clearly distinguishable by
future observations (see Figs.~\ref{fig:OurResultsonPlanck} and 
\ref{fig:OurResultsonPlanck2}).

From our analysis, we have found that the best choice of model in the power-law plateau
inflation family has the scalar potential \mbox{$V=V_0\varphi^2/(m^2+\varphi^2)$},
which is also a member of the shaft inflation family of models \cite{shaft}\footnote{%
However, in general, shaft inflation and power-law plateau inflation are different.}.

Such a potential was originally introduced by S-dual inflation in 
Ref.~\cite{Sdual}, where, however, the inflaton was non-canonically normalised
so the predicted value for $n_s$ was too large and incompatible with the Planck
data. Following Ref.~\cite{Sdual} but crucially considering canonically 
normalised fields (i.e. minimal K\"{a}hler potential) we have constructed a 
toy-model realisation in global and local supersymmetry for our preferred
power-law plateau inflation model.

All in all, the level of success of power-law plateau inflation, and the fact that it 
offers distinct and testable predictions make this a worthy candidate for 
primordial inflation, which may well be accommodated in a suitable theoretical 
framework, as our toy models suggest.

\paragraph{Acknowledgements}\leavevmode\\
%
%\noindent
CO is supported by the FST of Lancaster University. KD is supported (in part) by
the Lancaster-Manchester-Sheffield Consortium for Fundamental Physics under 
STFC grant: ST/L000520/1.

\newpage

%\begin{appendices}
%\chapter{Appendix A}

%\setcounter{table}{0} \renewcommand{\thetable}{A.\arabic{table}}

%\begin{longtable}{l|c|c|c|c|c}
%\caption{Maximum (non-integer) values n can take for each combination of $\alpha$, q, N whilst still remaining sub-Planckian and the corresponding values of $n_{s}$ and r}\\\hline
%	    \bfseries n & \bfseries q & \bfseries $\alpha$ & \bfseries N & \bfseries $n_{s}$ & \bfseries r \\\hline
%	    \endhead
%	    \hline
%	    \multicolumn{6}{r@{}}{continued \ldots}\\
%	    \endfoot
%	    \hline
%	    \endlastfoot
%	    \csvreader[head to column names]{FinalLimitCalcsforPaper.csv}{}% use head of csv as column names
 %   {\\\hline \n & \q & \alpha & \N & \ns & \r }% specify your coloumns here
  %  \label{tab:limits}
%	\end{longtable}
%\end{appendices}

\end{document}